\documentclass[aps,prl,twocolumn,groupedaddress,showpacs]{revtex4}

\usepackage{graphicx}
\usepackage{latexsym}

\begin{document}


\title{Measurement of the Spatial Correlation Function of Phase Fluctuating Bose-Einstein Condensates}


\author{D. Hellweg}
\email[]{hellweg@iqo.uni-hannover.de}
\author{L. Cacciapuoti}
\author{M. Kottke}
\author{T. Schulte}
\author{K. Sengstock$^1$}
\author{W. Ertmer}
\author{J.J. Arlt}
\affiliation{Institut f\"ur Quantenoptik, Universit\"at Hannover,
Welfengarten 1, 30167 Hannover, Germany\\$^1$Institut f\"ur
Laserphysik, Universit\"at Hamburg, Jungiusstra\ss e 2, 20355
Hamburg, Germany}


\date{\today}

\begin{abstract}
We measure the intensity correlation function of two interfering
spatially displaced copies of a phase fluctuating Bose-Einstein
Condensate (BEC). It is shown that this corresponds to a
measurement of the phase correlation properties of the initial
condensate. Analogous to the method used in the stellar
interferometer experiment of Hanbury Brown and Twiss, we use
spatial intensity correlations to determine the phase coherence
lengths of elongated BECs. We find good agreement with our
prediction of the correlation function and confirm the expected
coherence length.
\end{abstract}


\maketitle

Since the first realization of Bose-Einstein condensation in
dilute atomic gases their coherence properties have attracted
considerable theoretical and experimental interest. This interest
is due to the central role of the coherence properties for the
theoretical description and conceptional understanding of BECs and
their use as a source of coherent matter waves in many promising
applications.

Remarkable measurements demonstrated the phase coherence of
three-dimensional (3D) condensates well below the BEC transition
temperature $T_c$ \cite{Andrews1997b, Hagley1999b, Stenger1999b}
and even at finite temperature \cite{Bloch2000a}. However, low
dimensional systems show a qualitatively different behavior. In
particular, it has been predicted that one-dimensional and even
very elongated, three-dimensional BECs exhibit strong spatial and
temporal fluctuations of the phase while fluctuations in their
density distribution are suppressed
\cite{Petrov2000b,Petrov2001a}. Thus, the coherence properties are
significantly altered, resulting in a reduced coherence length
which can be much smaller than the condensate length. In this case
the degenerate sample is called a quasicondensate. This regime has
been subject of recent theoretical efforts based on the Bogoliubov
\cite{Mora2002a} and Popov \cite{Andersen2002b, Khawaja2002a}
theory to extend and generalize the description of BEC and to
calculate correlation functions \cite{Luxat2002a}. The phase
fluctuations are caused by thermal excitations of low energy axial
modes and thus depend strongly on the temperature and trapping
geometry. In particular, for very elongated BECs a nearly phase
coherent sample can only be achieved far below $T_c$. Phase
fluctuations were first observed using the formation of density
modulations during ballistic expansion of a condensate
\cite{Dettmer2001a, Hellweg2001a, Kreutzmann2002a}. In addition,
their effect on the momentum distribution has been demonstrated
using Bragg spectroscopy \cite{Richard2003a,Gerbier2002b}.
Nonequilibrium properties of these condensates have been studied
using a condensate focussing technique \cite{Shvarchuck2002a}.

\begin{figure}
\includegraphics[width=0.45\textwidth]{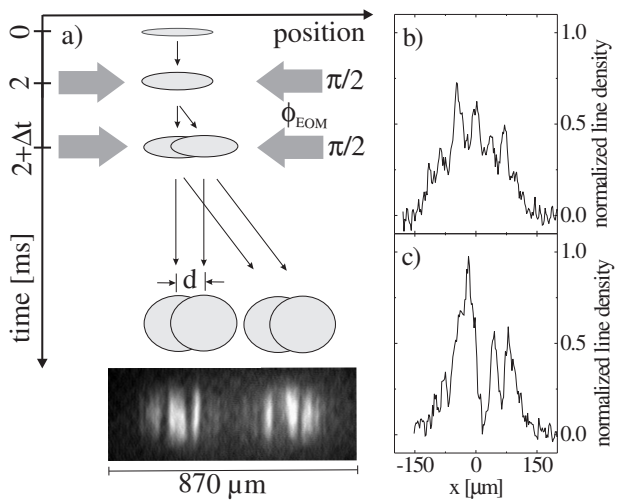}
\caption{(a) The interferometer is realized by two $\pi/2$ Bragg
diffraction pulses. We apply the first pulse after a ballistic
expansion time of $2\, $ms to reduce mean-field effects and atomic
scattering. The spatial displacement $d$ is determined by the time
$\Delta t$ between the pulses. Typical line density profiles are
shown for a phase coherence length of $L_\phi\approx25\,\mu$m with
$d=7\,\mu$m (b) and $d=35\,\mu$m (c). Only one of the two output
ports is displayed. \label{Fig1}}
\end{figure}

In this Letter we report on a direct measurement of the spatial
correlation function of phase fluctuating BECs. To measure the
phase correlation properties we interfere two copies of a BEC with
a spatial displacement $d$ (Fig.~\ref{Fig1}). The measured
interference pattern is determined by the phase pattern of the
original condensate and a global phase difference between the two
copies introduced by the interferometer. By varying the
displacement $d$ the first order correlation function can, in
principle, be measured. However, this method is very sensitive to
fluctuations in the global phase difference and the measurement is
further complicated by the statistical nature of phase
fluctuations. We show that the use of intensity correlations in
the interference pattern overcomes these problems and provides the
desired information about the phase correlations in the initial
condensate. The result of this measurement is described by the
spatial second order correlation function and yields the phase
coherence length of the BEC. In many respects our measurement is
closely related to the stellar interferometer of Hanbury Brown and
Twiss \cite{Hanbury1956a, Hanbury1956b}. They measured the
intensity of starlight falling onto two detectors and computed the
intensity correlations electronically as a function of the
detectors' distance. This measurement yields the transverse
coherence length of the stellar light and allowed them to
determine the stars' diameter. Unlike the Michelson stellar
interferometer, which uses first order correlations, atmospheric
fluctuations do not disturb this measurement. In the same way,
global phase fluctuations in the interferometer setup do not
disturb our measurement.

Our experiments were performed with $^{87}$Rb Bose-Einstein
condensates in the $|F\!=\!1, m_{F}\!=\!-1\rangle $ hyperfine
ground state. Further details of our experimental apparatus were
described previously \cite{Hellweg2001a}. The confining potential
was provided by a cloverleaf type magnetic trap with an axial
trapping frequency of $\omega_x=2\pi\times3.4\, $Hz and a radial
frequency adjusted between $\omega_r=2\pi\times300\, $Hz and
$\omega_r=2\pi\times380\, $Hz. The number of condensed atoms $N_0$
was varied between $4\times 10^4$ and $6\times 10^5$. To allow the
system to reach an equilibrium state we typically waited for $4\,
$s after obtaining BEC by evaporative cooling (with rf
"shielding") \footnote{After this time we do not observe any
quadrupole oscillations; we measured the thermalization time of
our condensates to be a few $100\, $ms.}. The interferometric
scheme is shown in Fig.~\ref{Fig1} and is based on two $\pi/2$
Bragg diffraction pulses. These pulses were produced by two
counterpropagating laser beams with a frequency difference set to
the two-photon resonance. They were detuned by about $3\, $GHz
from the atomic resonance to suppress spontaneous emission. The
pulse duration of $100 \, \mu$s was chosen long enough to avoid
higher order diffraction and sufficiently short to avoid any
sensitivity to the internal velocity distribution of the phase
fluctuating BECs. After a time-of-flight of 30 to 40 ms the two
output ports spatially separate and the atoms were detected by
resonant absorption imaging. We integrate the absorption images
along the radial direction of the condensate and subtract the
thermal background to obtain line density profiles.

Typical measured line density profiles are shown in
Fig.~\ref{Fig1}. These profiles provide a qualitative argument for
the use of intensity correlations as a particulary useful tool for
our measurements. When the displacement $d$ between the
overlapping clouds is chosen smaller than the phase coherence
length in the sample (Fig.~\ref{Fig1}b) regions with almost
identical phases are brought to overlap and the resulting
interference signal is rather smooth. If however $d$ is larger
than the phase coherence length, regions with substantially
different phases overlap, resulting in an irregular but high
contrast interference signal (Fig.~\ref{Fig1}c). In both cases an
average over many realizations and relative phases results in the
same smooth profile in the two output ports of the interferometer,
revealing no information about the coherence properties.
Nonetheless Fig.~\ref{Fig1}c clearly contains information about
the spatial coherence properties. An appropriate analysis of the
correlations in the density profile yields an intensity
correlation function that does not vanish in an averaging process
and contains the desired information about the coherence
properties.

We therefore start our analysis by calculating the spatial second
order correlation function for the case of phase fluctuating BECs.
In its most general definition \cite{Scully1997a} it is given by
\begin{equation}\label{gt2a}
g^{(2)}(x_1,x_2,x_3,x_4) =
\frac{\langle\hat\psi^\dagger(x_1)\hat\psi^\dagger(x_2)\hat\psi(x_3)\hat\psi(x_4)\rangle_T}
{\sqrt{\prod_{i=1}^4\langle\hat\psi^\dagger(x_i)\hat\psi(x_i)\rangle_T}},
\end{equation}
where $\langle\ldots\rangle_T$ denotes an average over an ensemble
at thermal equilibrium at temperature $T$. It contains the spatial
intensity correlation function
$g^{(2)}(x_1,x_2)=g^{(2)}(x_1,x_2,x_2,x_1)$ as a special case. For
3D condensates with repulsive interactions in elongated trapping
potentials density fluctuations are suppressed by the mean-field
potential \cite{Petrov2001a,Kreutzmann2002a,Richard2003a}.
\begin{figure}
\includegraphics[width=0.35\textwidth]{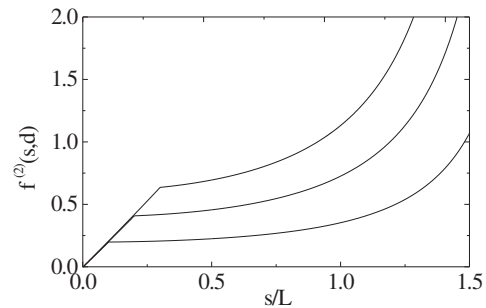}%
\caption{Numerical computation of the function $f^{(2)}(s,d)\equiv
f^{(2)}(\frac{-s-d}{2},\frac{s+d}{2},\frac{s-d}{2},\frac{-s+d}{2})$
for three different $d$. From top to bottom: $d/L=0.3,0.2,0.1$.
The particular choice of positions corresponds to the
experimentally relevant case.\label{Fig2}}
\end{figure}
Therefore the total field operator of the condensed atoms can be
written as $\hat\psi(x)=\sqrt{n_0(x)}\exp(i\hat\phi(x))$, where
$\hat\phi(x)$ is the operator of the phase \cite{Petrov2001a} and
$n_0(x)$ is the density in the Thomas-Fermi approximation. For
this field operator one can show that all higher order correlation
functions can be expressed as products of the first order
correlation function \cite{Cacciapuoti2003a}. Using the explicit
expression of the phase operator given by Petrov et
al.~\cite{Petrov2001a} we obtain the second order correlation
function:
\begin{equation}\label{gt2b}
g^{(2)}(x_1,x_2,x_3,x_4)=\exp\left[-\frac{1}{2l_\phi}f^{(2)}(x_1,x_2,x_3,x_4)\right],
\end{equation}
with
\begin{eqnarray}
\label{f2}
\lefteqn{\hspace{-0.5cm}f^{(2)}(x_1,x_2,x_3,x_4)=\frac{1}{8}\sum_{j}\frac{(j+2)(2j+3)}{j(j+3)(j+1)}
\left[P_j^{(1,1)}\left(\frac{x_1}{L}\right)\right.} \nonumber
\\&&\hspace{-0.5cm}\left.
+P_j^{(1,1)}\left(\frac{x_2}{L}\right)-P_j^{(1,1)}\left(\frac{x_3}{L}\right)
-P_j^{(1,1)}\left(\frac{x_4}{L}\right)\right]^2,
\end{eqnarray}
where $P_j^{(1,1)}$ are Jacobi polynomials, j is a positive
integer, $2L$ is the condensate length and
\begin{equation}
\label{deltal} l_\phi=\frac{L_\phi}{L}=\frac{15N_0(\hbar
\omega_x)^2}{32\mu k_B T}
\end{equation}
is the phase coherence length in the condensate center (in units
of $L$). Here $\mu$ denotes the chemical potential and $k_B$ the
Boltzmann constant. The function $f^{(2)}$ is shown in
Fig.~\ref{Fig2} and contains the functional form of phase
fluctuations in elongated condensates. All experimental parameters
are contained in the phase coherence length $l_\phi$.

Let us demonstrate how the interferometric sequence described
above can be used to measure $g^{(2)}$. In each output port of the
interferometer the superposition of two ballistically expanded
spatially displaced copies of the original wave function is
produced. The field operator of the atoms in one output port can
be expressed as
\begin{eqnarray}
\label{wavefct}
\hat\psi_f(x,d)=\frac{1}{2}\{\sqrt{n(x-d/2)}e^{i\hat\phi^{+}(x-d/2)}
\nonumber \\+
\sqrt{n(x+d/2)}e^{i\hat\phi^{-}(x+d/2)}e^{i\phi_{\mbox{\tiny
rel}}}\},
\end{eqnarray}
with $\hat\phi^{\pm}(x)=\hat\phi(x)\pm \beta x+\alpha x^2$
\footnote{The origin of the x-axis has been set to the center of
the two overlapping clouds.}. The linear term results from the
mean-field repulsion between the copies of the condensate
\cite{Simsarian2000a} and the quadratic term originates from the
self-similar expansion \cite{Castin1996a,Kagan1996a}. The
self-similar expanded density distribution $n(x)$ differs only
slightly from the initial distribution because the increase of
axial size during expansion is small for very elongated
condensates (about 1\% for our experimental conditions). The
relative global phase between the two overlapping clouds
$\phi_{\mbox{\scriptsize rel}}=\delta_{12}\Delta
t+\delta\phi_{\mbox{\scriptsize eff}}$ is determined by the Bragg
diffraction process, where $\delta\phi_{\mbox{\scriptsize eff}}$
is the change in the relative phase of the Bragg beams between the
two pulses and $\delta_{12}$ is the detuning from the two-photon
resonance. In our experiment $\delta\phi_{\mbox{\scriptsize eff}}$
can be controlled by using an electro-optical modulator
\cite{Torii2000a}, $\delta_{12}$ depends on the frequency
difference of the Bragg beams and the axial release velocity of
the condensate. Note that even a small change in the release
velocity of $0.033\, \mu$m/ms leads to a phase change of $\pi/2$
for a time $\Delta t=3\, $ms between the Bragg pulses
\footnote{Compared to previous autocorrelation measurements with
BECs \cite{Simsarian2000a} our signal is highly sensitive to such
fluctuations because of the long $\Delta t$ needed to reach a
sufficient displacement.}. The insensitivity to such randomly
varying global phases is a major advantage of the intensity
correlation method.

In analogy to the definition of the correlation coefficient we
define a normalized intensity correlation function
\begin{equation}
\label{corrdef} \gamma^{(2)}_f(x_1,x_2,d)=\frac{\langle(\hat
I_1-\langle\hat I_1\rangle)(\hat I_2-\langle\hat
I_2\rangle)\rangle} {\sqrt{\langle(\hat I_1-\langle\hat
I_1\rangle)^2\rangle}\sqrt{\langle(\hat I_2-\langle\hat
I_2\rangle)^2}\rangle},
\end{equation}
with the intensity operator $\hat
I_{1,2}=\hat\psi_f^{\dagger}(x_{1,2},d)\hat\psi_f(x_{1,2},d)$.
Here the averages are taken over an ensemble in thermal
equilibrium and over all relative phases. The possible values of
$\gamma^{(2)}_f$ range between +1 (perfect correlation) and -1
(perfect anticorrelation); if $\hat I_1$ and $\hat I_2$ are
uncorrelated $\gamma^{(2)}_f=0$. By substituting
Eq.~(\ref{wavefct}) into Eq.~(\ref{corrdef})  we obtain
\begin{eqnarray}
\label{corrcalc} \lefteqn{\gamma^{(2)}_{f}(x_1,x_2,d)=\cos[2(x_1-x_2)(\alpha d-\beta)]}\nonumber\\
&&\hspace{-0.5cm}\times\exp\left[-\frac{1}{2l_\phi}f^{(2)}(x_1-\frac{d}{2},x_2+
\frac{d}{2},x_2-\frac{d}{2},x_1+\frac{d}{2})\right].
\end{eqnarray}
The normalized intensity correlation function is the product of
two contributions. A cosine resulting from the self-similar
expansion and mean-field repulsion and an exponentially decaying
term containing the influence of the phase fluctuations. The decay
constant of this function is the phase coherence length.
Comparison with Eq.~(\ref{gt2b}) shows that a measurement of
$\gamma^{(2)}_{f}$ is equivalent to a measurement of the second
order correlation function of the trapped condensate.

\begin{figure}
\includegraphics[width=0.45\textwidth]{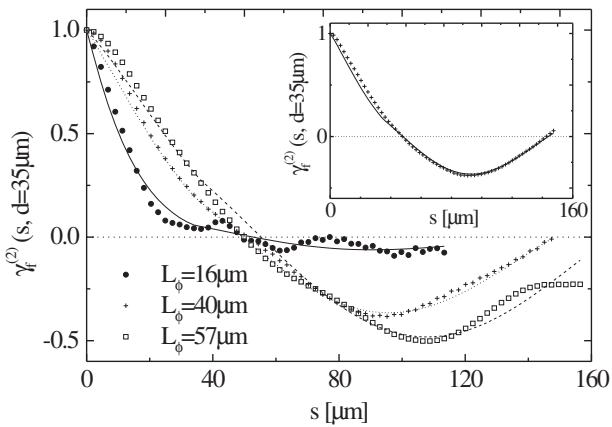}%
\caption{Measured correlation functions (points) compared to
fitted theoretical curves (lines). Each measured function is based
on $\approx25$ realizations and plotted up to $s=0.8L$.
Neighboring points are not independent since they are based on a
common set of experimental realizations. Inset: Numerical
simulation including the phase and density evolution (points) and
the analytical function Eq.~(\ref{corrcalc}) (line), both for the
parameters of the $L_\phi=40\,\mu$m curve. \label{Fig3}}
\end{figure}

The measurement of $\gamma^{(2)}_{f}$ is performed in the
following way: For a given trap configuration, evaporative cooling
ramp and fixed displacement $d$, a series of measurements is
recorded by scanning the relative phase of the Bragg beams with
the electro-optical modulator in small steps between 0 and $2\pi$.
This insures that the global phase $\phi_{\mbox{\scriptsize rel}}$
contains all values with equal probability. We experimentally
determine $\gamma^{(2)}_{f}$ analogous to Eq.~(\ref{corrdef}). The
ensemble averages $\langle\hat I(x)\rangle$ are obtained by
averaging all interference patterns $I(x)$ recorded in a series.
Then the quantity $I(x)-\langle\hat I(x)\rangle$ can be determined
for each realization. The average of these values according to
Eq.~(\ref{corrdef}) yields $\gamma^{(2)}_{f}$. To simplify the
analysis we evaluate $\gamma^{(2)}_{f}$ at symmetric positions
around the center of the interference pattern such that
$x_1=-x_2=s/2$. Then
$\gamma^{(2)}_f(s,d)=\gamma^{(2)}_f(-s/2,s/2,d)$ can be expressed
as a function of $s$ for a given displacement $d$. Typical results
are shown in Fig.~\ref{Fig3}, clearly displaying the functional
form of a damped cosine.

To extract quantitative results we fit the measured function with
the theoretical one given in Eq.~(\ref{corrcalc}). The fit
contains only $l_\phi$ and the frequency of the cosine as free
parameters. Figure \ref{Fig3} compares measured correlation
functions with the corresponding fits, confirming the excellent
agreement with the expected functional form. Although the phase
coherence length also depends on the trapping potential and the
temperature, the data sets shown in Fig.~\ref{Fig3} differ mainly
due to the number of condensed atoms, which was set to $4.4\times
10^4 (\bullet)$, $2.9\times 10^5 (\mbox{\scriptsize +})$ and
$5.0\times 10^5 (\Box)$. Except for the smallest atom number a
minimum is clearly visible and unambiguously defines the frequency
of the cosine \footnote{Note that the measured frequency of the
cosine is in good agreement with the prediction of $\alpha$ and
$\beta$.}. The damping of this oscillation yields the phase
coherence length. As expected (see Eq.~(\ref{deltal})), the
damping is stronger for small $N_0$. In case of the smallest $N_0$
no oscillation is visible, indicating a significant decrease of
the phase coherence length.

\begin{figure}
\includegraphics[width=0.4\textwidth]{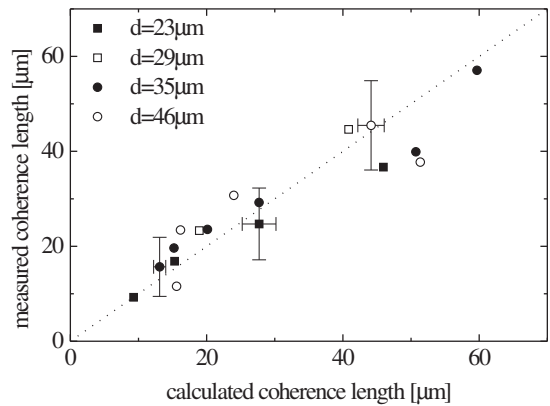}%
\caption{Measured phase coherence length compared to the
theoretically expected one given by Eq.~(\ref{deltal}). The dotted
line $f(x)=x$ is a guide to the eye. The error bars indicate
statistical errors. Systematic uncertainties are 26\% and 15\% for
the calculated and measured $L_\phi$, respectively. \label{Fig4}}
\end{figure}

We have performed such measurements for a large variety of atom
numbers and temperatures. The good agreement between the measured
phase coherence length obtained from the fit and the theoretically
predicted one is shown in Fig.~\ref{Fig4}. By varying the
displacement $d$ we have confirmed that the measured phase
coherence lengths are independent of this parameter. In all cases
the phase coherence length was much smaller than the condensate
length, which ranged from $280\, \mu$m up to $420\, \mu$m,
i.e.~our measurements were performed in the quasicondensate
regime.

So far we have neglected the evolution of the phase fluctuations
during ballistic expansion. This evolution leads to a change of
the original phase pattern and the appearance of density
modulations. Using the full evolution of the wave function
\cite{Cacciapuoti2003a} we have calculated the expected phase
change during time-of-flight to be less than $\pi/10$ for our
parameters. To evaluate the influence of density modulations on
our measurements we have performed numerical simulations including
the phase and density evolution. The inset in Fig.~\ref{Fig3}
compares a numerical result with Eq.~(\ref{corrcalc}). The
excellent agreement demonstrates that this evolution can be
neglected for our intensity correlation measurements and justifies
the use of Eq.~(\ref{corrcalc}) to extract the phase correlation
properties.

In conclusion, we have demonstrated a new interferometric method
which allows us to measure the spatial correlation function of
phase fluctuating BECs. First the second order correlation
function was calculated and it was shown that this function can be
measured using intensity correlations in the interference pattern.
Our measurements were then compared with these results, confirming
both, the expected functional form of the correlation function and
the phase coherence length of the sample. We have confirmed that
this technique is insensitive to fluctuations of the relative
global phase during the interferometric measurement sequence.

 We thank L. Santos for valuable discussions
and calculations on the expansion of phase fluctuating BECs. This
work is supported by the {\it Deutsche Forschungsgemeinschaft}
within the SFB\,407.

\bibliography{Hellweg}

\end{document}